# EMPLOYING HYBRID DEEP NEURAL NETWORKS ON DARI SPEECH

Jawid Ahmad Baktash[1] and Mursal Dawodi[2]

LIA, Avignon University
Avignon, France
[1]jawid.baktash1989@gmail.com, [2]mursal.dawodi@gmail.com

## ABSTRACT

*This paper is an extension of our previous conference paper. In recent years, there has been a growing interest among researchers in developing and improving speech recognition systems to facilitate and enhance human-computer interaction. Today, Automatic Speech Recognition (ASR) systems have become ubiquitous, used in everything from games to translation systems, robots, and more. However, much research is still needed on speech recognition systems for low-resource languages. This article focuses on the recognition of individual words in the Dari language using the Mel-frequency cepstral coefficients (MFCCs) feature extraction method and three different deep neural network models: Convolutional Neural Network (CNN), Recurrent Neural Network (RNN), and Multilayer Perceptron (MLP), as well as two hybrid models combining CNN and RNN. We evaluate these models using an isolated Dari word corpus that we have created, consisting of 1000 utterances for 20 short Dari terms. Our study achieved an impressive average accuracy of 98.365%.*

## KEYWORDS

*Dari, deep neural network, speech recognition, recurrent neural network, multilayer perceptron, convolutional neural network*

## 1. INTRODUCTION

Humans communicate with each other using speech, and Automatic Speech Recognition (ASR) tools have greatly facilitated human-machine interaction with advancements in technology and the widespread use of smartphones. ASR systems have numerous applications, including machine-human dialogue systems, robotics, translators, and more. However, research on ASR for many regional and Asian languages still faces many challenges, such as a lack of resources and different dialects.

Dari, one of Afghanistan's official languages, is spoken as a first or second language by about 32 million people worldwide. Although many Dari speakers live outside Afghanistan, they often face communication barriers due to their limited access to modern technology devices and applications. Furthermore, research on ASR for Dari and other regional languages is still in its infancy, with only a handful of studies to date.

This article aims to develop a more robust and less error-prone ASR system for Dari by employing three different deep learning techniques, including Multilayer Perceptron (MLP), Convolutional Neural Network (CNN), Long Short-term Memory (LSTM), a hybrid model of CNN and LSTM, and a hybrid model of CNN and Bidirectional LSTM (BLSTM), to develop Dari isolated words ASR. We then compare the performance of the different models to evaluate their efficiency. This study is novel in the field of Dari speech recognition as it compares the effectiveness of five state-of-the-art deep learning approaches.

To the best of our knowledge, no published article has compared the performance of different models on Dari ASR. The primary challenge in developing an ASR system for Dari is the lack of available datasets and the vast variety of dialects. We address this challenge by developing and using our built-in isolated Dari words corpus that consists of 1000 utterances for 20 short Dari terms.

Authors by Dawodi et al. [3] established an ASR system for Dari using MFCC for feature extraction and CNN for word classification, achieving an average accuracy of 88.2%. However, the current study builds upon their work by implementing three additional deep learning techniques and achieves more than 10% higher accuracy. The next section will discuss some related works in this field, and Section 3 will briefly introduce the corpus structure. Subsequently, sections 4 to 6 will describe an overview of Dari speech recognition, including MFCC feature extraction and deep neural network models. The result and discussion on this work will be illustrated in sections 7. Finally, section 8 will conclude this article and report future work. Additionally, the authors will clarify that this paper is an extension of their previous conference paper.

## 2. RELATED WORKS

Several studies have used machine learning and NLP techniques for speech recognition tasks, with recent focus on deep learning techniques for developing ASR systems. However, most studies have focused on non-regional languages such as English. Mitra et al. [8] proposed a hybrid convolution neural network (HCNN) for modeling acoustic and articular spaces. The HCNN model showed better performance than CNN/DNN, with a lower word error rate. Similarly, Grozdic and Jovicic [6] developed a new framework based on the automatic deep decoder encoder coefficient and Teager energy storage coefficients. They showed that Teager energy-based cepstral properties are more powerful than MFCC and GMM-HMM in describing whispers, with 31% higher accuracy in the whisper scenario and 92.81% word recognition rate.

During the recent decade, some studies focused on more regional languages. An ASR for the Panjabi language was developed by Dua et al. [4] that used MFCC for features extraction and HTK toolkit for recognition. They prepared the dataset containing 115 Panjabi terms by the utterance of 8 Punjabi native speakers. The overall mean accuracy achieved was between 94%–96%. Bakht Zada and Rahimullah [17] developed a Pashto isolated digit ASR to detect Pashto digits from zero to nine. They used MFCC to extract features and CNN to classify digits. The utilized dataset consisted of 50 sounds for each number. Their proposed model contained four convolutional layers, followed by ReLU and max-pooling layers. Subsequently, they obtained an average accuracy of 84.17%. Dahl et al. [2] developed a novel context-dependent model for speech recognition using deep neural network techniques. They proved that their proposed model is superior to previous context-dependent methods. Similarly, Abdel-Hamid [1] used CNN in the speech recognition context and proved its efficiency in decreasing the error-rate and increasing robustness. Graves et al. [5] proposed a hybrid model that involved bidirectional Long Short-term Memory (LSTM) RNNs and weight noise. They evaluate their model on the TIMIT phoneme recognition benchmark. As a result, the error rate decreased by up to 17.7%.

Few recently published articles focus on ASR for the Persian language using different models. Sameti et al. [12] implemented a Persian continuous speech recognition system. They used MFCC with some modification to learn features of speech signals and model-based techniques, speech enhancement approaches like spectral subtraction, and Wiener filtering to gain desirable robustness. Likewise, Hasanabadi et al. [7] used a database containing Persian isolated words and developed Persian ASR in 2008. Finally, they create a wheeled mobile robot to navigate using Persian spoken commands. They investigated simple Fast Fourier Transform (FFT) to catch attributes and MLP to classify patterns. S. Malekzadeh et al. used MFCC for extracting features from Persian speeches and MLP for detecting vowel and consonant characters. The used dataset involved 20 categories of acoustics from utterances of 10 people. The proposed model demonstrated 61% - 87% conversion accuracy. S. Malekzadeh utilized a deep MLP deep model to recognize Persian sounds to improve voice signal processing. Similarly, [9] established a Persian letter and word to sound system. They utilized rule-based for the first layer and MLP for the other layers of the network model. The average accuracy of 60.7% and 83% were obtained for letter and word predictions respectively. Recently, Veisi and Mani [15] used a hybrid model of deep belief network (DBN) for extracting features and DBLSTM with Connectionist Temporal Classification (CTC) output layer to create the acoustic model. The study indicates that DBLSTM provides higher accuracy in Persian phoneme recognition compared to the traditional models.

This paper presents the design of an ASR for Dari isolated words. The works presented in this paper is based on three different deep learning techniques.

## 3. DARI WORD CORPUS

In this research, we utilized a dataset that we had previously created [3]. The dataset consists of 1000 sounds, representing 20 short Dari terms spoken by 20 Dari native speakers of both genders and different dialects. The speakers recorded the utterances in a noise-free environment at their homes and offices using a smartphone audio recorder. The recorded sounds were then transferred to a PC using a USB cable, and we used Adobe Audition software to remove background noise and limit the length of each file to one second. Finally, we saved all audio files in the .wav extension format, with each word saved in a separate file bearing its name. All utterance files related to a single term are located within the same folder; for instance, all audio records associated with the "Salaam" term from all speakers are stored in the "fold 20" folder. As a result, there are 20 separate folders, each containing the utterances relevant to a single term. We ended up with 1000 usable utterances, as some of the files were corrupted.

## 4. DARI ASR SYSTEM

This study represents a state-of-the-art approach to Dari ASR. The proposed system uses audio signals as inputs, MFCC to extract features from voice signals, and five separate deep learning models to recognize utterances. Additionally, we divided the dataset into training and testing sets with different ratios to evaluate the performance of the architectures. Each method was trained and tested, and the entire models were able to accurately detect the term and output it.

Feature extraction is a crucial step in analyzing data and identifying relationships between various objects, such as audio signals. Models are unable to recognize utterances directly, which is why speech signals need to be transformed into an understandable format. Additionally, features must be extracted from the raw waveform to minimize variability in speech signals and produce input that is appropriate for modeling. This study utilizes MFCC, which is the most dominant method for extracting key attributes [16]. The MFCC of a signal is a set of features that represents the entire shape of a spectral envelope.

## 5. NEURAL NETWORK MODELS

In our conference paper we used three deep neural network models. Our deep CNN model consisted of four layers, each containing 2D convolutional layers followed by max-pooling layers except for the last layer, which uses global-average-pooling. Subsequently, it has a fully connected layer in the last as an output layer. The input of the first convolution layer is a 2D tensor of 40174 (representing MFCC features) which is convolved with 16 convolutional kernels. The next convolution layer is convolved with 32 filters. We doubled the number of kernels in the subsequent layers. The output of the preceding layer is the input of the succeeding layer. Every convolutional filter generates a feature map based on input. As an example, the first layer generates 16 feature maps while the last layer provides 128 feature maps. Each convolution layer convolves with 2 filters. We used the linear activation function to overcome the gradient vanishing problem [8]. We used ReLU, which is a well-known activation function in this study. In the next step, max pooling is applied to the features to reduce the sampling of the feature map by dividing the feature map into a small rectangular area, usually called the window, which does not overlap with each other (Ide & Kurita, 2017). Studies have shown that max pooling is more effective than other pooling types (Ide & Kurita, 2017). Hence, we used max pooling with a pool size of 2 instead of other pooling layers for the first three blocks. Usually, the performance of a model is affected by noise samples during training. Therefore, dropout is applied to address this problem and prevent a neural network from overfitting [13]. We tried various probability values (in the range of 0 and 2.5) for dropout. As a result, the most effective probability value is 2.0, which enhances the accuracy and minimizes the loss due to the limitation of our dataset for each term. The output size is the same as the total number of classes, which is 20 classes each related to a certain word. We examined the impact of different epochs and batch size on this model during training. Consequently, the optimal result was obtained with 80 epochs with a batch size of 84.

The MLP model consists of two hidden layers each is followed by the ReLU activation function. Every single layer, except the output layer, has 300 neurons. Afterwards, a dropout of 2.0 is applied to decrease the likelihood of overfitting on the training data by randomly excluding nodes from each cycle. The input shape for this model is a one-dimensional feature vector that contains 40 features. The output layer contains 20 nodes, each representing a class, and is followed by the SoftMax activation function, like in the CNN model. The goal behind selecting SoftMax as the activation function for the last layer is that it sums up the output to 1. Finally, the model was trained with a batch size of 32 and a total number of 108,620 parameters.

The structure of the implemented RNN model is a sequential model. It consists of one Long Short-Term Memory (LSTM) architecture block with a size of 64, which is followed by a dropout layer with a rate of 0.2. The LSTM model uses the Tanh activation function to activate the cell mode and the sigmoid activation function for the node output. The next layer is the Flatten layer, which collapses the spatial dimensions of the input into the channel dimension. The output of the Flatten layer is then passed to the output layer, which consists of 20 nodes. Finally, the output layer is followed by the SoftMax activation function, similar to the MLP and CNN models. The best performance was obtained using 100 epochs with a batch size of 64. We trained each model with several different epochs and batch sizes, and the most appropriate value that resulted in the highest accuracy was selected.

## 6. HYBRID MODELS

We implemented two distinct hybrid models to evaluate the performance of hybrid deep learning techniques in our Dari ASR system. The first model is composed of four 2-dimensional convolution hidden layers followed by a single LSTM hidden layer. Additionally, it contains a fully connected layer and an output layer at the end. The second hybrid model is like the first one. However, we substitute the LSTM with the Bidirectional LSTM technique and flattened the output of LSTM before feeding it to the output layer. Each hidden layer is followed by the ReLU activation function. Additionally, a 2D max-pooling layer is associated after every convolutional 2D layer like the deep CNN structure. The capacity of LSTM is 64 in both models. We manually fine-tuned the number of layers, dropout rate, epochs, and batch-size to reach the optimal value. As a result, a 20% dropout is implied after a max-pooling layer. Consequently, the models were separately trained in 100 epochs using a batch size of 64 to achieve the best result.

## 7. RESULTS AND EXPERIMENTS

We have developed an ASR system for Dari language to recognize 20 isolated words. To build the system, we used a dataset from our previous work which contains 50 utterances for each word. We applied MFCC for feature extraction and used MLP, CNN, LSTM, CNN+LSTM, and CNN+BLSTM for speech classification. The models were trained and evaluated using 10-fold cross-validation, categorical cross-entropy as loss function, accuracy as metric, and Adam optimizer. The implementation was done in Python 3.6.9 on a Windows 10 computer with an Intel core i7(TM) 2.80 GHz 2.90 GHz processor. Feature extraction was performed using Librosa version 0.8.0, while TensorFlow version 2.3.1 and Keras version 2.4.3 were used for implementing the models.

Table 1 shows the average accuracy and loss during training and testing for each model. The CNN+BLSTM model achieved the highest average testing accuracy of 98.365%, followed by CNN+LSTM and CNN with 98.184% and 98.037%, respectively. Although combining CNN with LSTM or BLSTM had minimal effects on accuracy and loss, it increased the complexity and runtime of the approach. Therefore, we recommend using the CNN+BLSTM approach for minor improvements in accuracy or loss dimensions. LSTM, although suitable for many NLP applications due to its nonlinear nature, showed comparable average testing accuracy to MLP due to the small size and simplicity of the dataset. However, MLP had the highest average testing loss value among all models.

Tables 2-6 show the sum of 10 confusion matrices for each model. The LSTM model had the lowest average training loss and highest training accuracy but lower average testing accuracy. CNN had a maximum accuracy of 99.63% in fold 8, but this dropped to 95.63% in fold 6. The CNN+BLSTM model achieved a maximum accuracy of 99.27% in folds 4, 6, and 7, and a minimum accuracy of 97.45% in fold 9.

Training and testing the MLP model for one fold took about 11 seconds, while it took about 2 minutes and 32 seconds to train and test the LSTM architecture for 100 iterations. The CNN model took much longer (approximately 14 minutes and 12 seconds for 80 cycles) due to its complexity. The execution time of hybrid models CNN+LSTM and CNN+BLSTM was even longer, with approximately 19 minutes and 15 minutes and 45 seconds spent on training and testing the models in 100 epochs, respectively.

Table 1. Training and testing accuracy and loss

| Model | Average Training Accuracy | Average Training Loss | Average Testing Accuracy | Average Testing Loss |
|---|---|---|---|---|
| **MLP** | 99.73 | 0.028 | 97.02 | 0.129 |
| **CNN** | 99.93 | 0.018 | 98.037 | 0.105 |
| **LSTM** | 100 | 0 | 97.712 | 0.09 |
| **CNN+ LSTM** | 99.95 | 0.006 | 98.184 | 0.079 |
| **CNN+ BLSTM** | 99.98 | 0.006 | 98.365 | 0.07 |

In this study, we aimed to identify the best performing models by modifying various parameters such as dense layers, pooling layers, dropouts, testing and training iterations, batch size, and kernel size. Based on the results obtained, we selected the optimal options for our research. Among the five models evaluated, the CNN+BLSTM model achieved the most impressive average accuracy results of 99.98% during training and 98.365% during testing, surpassing even the LSTM model which showed better average accuracy during training (100%) but lower accuracy during testing, as shown in table 1.

Our research outperformed previous studies on Dari isolated word speech recognition. For instance, Dawodi et al. [3] employed MFCC and CNN to develop a Dari ASR for isolated words, but our model achieved over 10% higher accuracy on the test data. Moreover, our methods also exhibited lower evaluation loss and higher accuracy compared to other recent ASR studies. Table 7 presents a comparison of our research with some of the latest and innovative ASR systems.

## 8. CONCLUSIONS

This paper is an extension of our conference paper and presents a study on the effectiveness of five deep neural network models, namely MLP, CNN, RNN, CNN+LSTM, and CNN_BLTSM, for recognizing isolated Dari words through audio signal features extracted using MFCC. The results showed that CNN and hybrid CNN models outperformed the other models in Dari ASR. While this study provides a good starting point for Dari natural language processing, further research is needed to develop more accurate and continuous Dari ASR models using hybrid models and novel techniques. However, this would require a larger and more diverse dataset.

Table 2. Confusion-matrix MLP model

| MLP | 1 | 2 | 3 | 4 | 5 | 6 | 7 | 8 | 9 | 10 | 11 | 12 | 13 | 14 | 15 | 16 | 17 | 18 | 19 | 20 |
|---|---|---|---|---|---|---|---|---|---|---|---|---|---|---|---|---|---|---|---|---|
| 1 | 169 | 0.5 | 0 | 0 | 0 | 0 | 0 | 0 | 0 | 0 | 2 | 0 | 0 | 0 | 0 | 0 | 0 | 0 | 0 | 0 |
| 2 | 0 | 93 | 0 | 0 | 0 | 0 | 0 | 0 | 0 | 0 | 0 | 0 | 0 | 0 | 0 | 0 | 0 | 0 | 0 | 0 |
| 3 | 5 | 0 | 220 | 0 | 0 | 0 | 1 | 0 | 0 | 0 | 0 | 0 | 0 | 0 | 0 | 0 | 0 | 0 | 0 | 0 |
| 4 | 0 | 0 | 0 | 135 | 0 | 0 | 2 | 0 | 0 | 0 | 0 | 0 | 0 | 0 | 0 | 0 | 0 | 3 | 0 | 0 |
| 5 | 0 | 0 | 0 | 0 | 131 | 0 | 0 | 0 | 0 | 0 | 0 | 0 | 0 | 0 | 0 | 0 | 0 | 0 | 0 | 0 |
| 6 | 0 | 0 | 0 | 0 | 0 | 135 | 0 | 0 | 0 | 0 | 0 | 0 | 0 | 0 | 0 | 0 | 0 | 0 | 0 | 0 |
| 7 | 0 | 0 | 0 | 0 | 0 | 0 | 227 | 0 | 0 | 0 | 0 | 0 | 0 | 0 | 0 | 0 | 0 | 0 | 1 | 0 |
| 8 | 0 | 0 | 2 | 0 | 0 | 0 | 0 | 108 | 0 | 0 | 0 | 0 | 2 | 4 | 0 | 0 | 2 | 0 | 0 | 0 |
| 9 | 0 | 0 | 0 | 0 | 0 | 0 | 0 | 0 | 132 | 0 | 0 | 0 | 0 | 0 | 0 | 0 | 0 | 0 | 0 | 0 |
| 10 | 0 | 2 | 0 | 0 | 0 | 2 | 0 | 1 | 0 | 120 | 1 | 0 | 3 | 0 | 0 | 5 | 0 | 0 | 0 | 0 |
| 11 | 0 | 0 | 0 | 0 | 0 | 0 | 0 | 0 | 0 | 0 | 88 | 0 | 2 | 0 | 0 | 0 | 0 | 0 | 2 | 0 |
| 12 | 4 | 0 | 0 | 0 | 0 | 0 | 0 | 0 | 0 | 0 | 0 | 84 | 0 | 2 | 0 | 0 | 0 | 0 | 0 | 0 |
| 13 | 0 | 0 | 0 | 0 | 0 | 0 | 1 | 2 | 0 | 3 | 2 | 0 | 162 | 0 | 0 | 0 | 0 | 0 | 0 | 2 |

| | 1 | 2 | 3 | 4 | 5 | 6 | 7 | 8 | 9 | 10 | 11 | 12 | 13 | 14 | 15 | 16 | 17 | 18 | 19 | 20 |
|---|---|---|---|---|---|---|---|---|---|---|---|---|---|---|---|---|---|---|---|---|
| 14 | 0 | 0 | 2 | 0 | 0 | 0 | 0 | 0 | 0 | 0 | 0 | 0 | 0 | 112 | 0 | 0 | 0 | 0 | 0 | 0 |
| 15 | 0 | 0 | 0 | 0 | 0 | 0 | 0 | 0 | 0 | 0 | 0 | 0 | 0 | 0 | 152 | 0 | 0 | 0 | 0 | 0 |
| 16 | 0 | 0 | 0 | 0 | 0 | 2 | 3 | 0 | 0 | 1 | 2 | 0 | 1 | 0 | 0 | 88 | 0 | 1 | 0 | 0 |
| 17 | 0 | 0 | 0 | 0 | 0 | 0 | 0 | 0 | 0 | 0 | 0 | 0 | 0 | 0 | 0 | 0 | 140 | 0 | 0 | 0 |
| 18 | 0 | 0 | 0 | 0 | 0 | 0 | 0 | 0 | 0 | 0 | 0 | 0 | 0 | 0 | 0 | 0 | 0 | 136 | 0 | 0 |
| 19 | 0 | 0 | 0 | 0 | 0 | 0 | 0 | 0 | 0 | 0 | 0 | 0 | 0 | 0 | 0 | 0 | 0 | 0 | 128 | 0 |
| 20 | 2 | 0 | 0 | 0 | 0 | 0 | 1 | 0 | 0 | 2 | 0 | 0 | 1 | 0 | 0 | 0 | 0 | 0 | 1 | 111 |

Table 3. Confusion-matrix LSTM model

| LSTM | 1 | 2 | 3 | 4 | 5 | 6 | 7 | 8 | 9 | 10 | 11 | 12 | 13 | 14 | 15 | 16 | 17 | 18 | 19 | 20 |
|---|---|---|---|---|---|---|---|---|---|---|---|---|---|---|---|---|---|---|---|---|
| 1 | 174 | 0 | 0 | 0 | 0 | 0 | 2 | 0 | 0 | 0 | 0 | 0 | 0 | 0 | 0 | 0 | 0 | 0 | 0 | 0 |
| 2 | 0 | 90 | 0 | 0 | 0 | 0 | 0 | 3 | 0 | 0 | 0 | 0 | 0 | 0 | 0 | 0 | 0 | 0 | 0 | 0 |
| 3 | 0 | 0 | 218 | 2 | 0 | 0 | 0 | 0 | 0 | 0 | 0 | 0 | 0 | 4 | 2 | 0 | 0 | 0 | 0 | 0 |
| 4 | 0 | 0 | 0 | 136 | 0 | 0 | 2 | 0 | 0 | 0 | 0 | 0 | 0 | 0 | 0 | 2 | 0 | 0 | 0 | 0 |
| 5 | 0 | 0 | 0 | 0 | 131 | 0 | 0 | 0 | 0 | 0 | 0 | 0 | 0 | 0 | 0 | 0 | 0 | 0 | 0 | 0 |
| 6 | 0 | 0 | 0 | 0 | 0 | 135 | 0 | 0 | 0 | 0 | 0 | 0 | 0 | 0 | 0 | 0 | 0 | 0 | 0 | 0 |
| 7 | 0 | 0 | 0 | 0 | 0 | 0 | 228 | 0 | 0 | 0 | 0 | 0 | 0 | 0 | 0 | 0 | 0 | 0 | 0 | 0 |
| 8 | 0 | 0 | 0 | 0 | 0 | 0 | 0 | 114 | 0 | 0 | 0 | 0 | 2 | 2 | 0 | 0 | 0 | 0 | 0 | 0 |
| 9 | 0 | 0 | 0 | 0 | 0 | 0 | 0 | 0 | 132 | 0 | 0 | 0 | 0 | 0 | 0 | 0 | 0 | 0 | 0 | 0 |
| 10 | 2 | 0 | 0 | 0 | 0 | 0 | 0 | 2 | 0 | 122 | 0 | 0 | 2 | 2 | 0 | 0 | 2 | 0 | 0 | 2 |
| 11 | 0 | 0 | 0 | 0 | 0 | 0 | 0 | 0 | 0 | 0 | 92 | 0 | 0 | 0 | 0 | 0 | 0 | 0 | 0 | 0 |
| 12 | 0 | 0 | 0 | 0 | 0 | 0 | 0 | 0 | 0 | 0 | 0 | 90 | 0 | 0 | 0 | 0 | 0 | 0 | 0 | 0 |
| 13 | 0 | 0 | 0 | 0 | 0 | 0 | 0 | 0 | 2 | 2 | 0 | 0 | 168 | 0 | 0 | 0 | 0 | 0 | 0 | 0 |
| 14 | 0 | 2 | 0 | 0 | 0 | 0 | 0 | 2 | 0 | 0 | 0 | 0 | 2 | 108 | 0 | 0 | 0 | 0 | 0 | 0 |
| 15 | 0 | 0 | 0 | 0 | 0 | 0 | 0 | 0 | 0 | 0 | 0 | 0 | 0 | 0 | 152 | 0 | 0 | 0 | 0 | 0 |
| 16 | 0 | 0 | 0 | 0 | 0 | 0 | 0 | 0 | 0 | 2 | 2 | 0 | 0 | 0 | 0 | 94 | 0 | 0 | 0 | 0 |
| 17 | 0 | 0 | 0 | 0 | 0 | 0 | 0 | 0 | 0 | 0 | 0 | 0 | 0 | 0 | 0 | 0 | 140 | 0 | 0 | 0 |
| 18 | 0 | 0 | 0 | 0 | 0 | 0 | 0 | 0 | 0 | 0 | 0 | 0 | 0 | 0 | 0 | 0 | 0 | 136 | 0 | 0 |
| 19 | 0 | 0 | 0 | 0 | 0 | 0 | 0 | 0 | 0 | 0 | 0 | 0 | 0 | 0 | 0 | 0 | 0 | 0 | 128 | 0 |
| 20 | 0 | 0 | 4 | 0 | 0 | 0 | 2 | 0 | 0 | 4 | 2 | 0 | 2 | 0 | 2 | 0 | 0 | 0 | 0 | 102 |

Table 4. Confusion-matrix CNN model

| CNN | 1 | 2 | 3 | 4 | 5 | 6 | 7 | 8 | 9 | 10 | 11 | 12 | 13 | 14 | 15 | 16 | 17 | 18 | 19 | 20 |
|---|---|---|---|---|---|---|---|---|---|---|---|---|---|---|---|---|---|---|---|---|
| 1 | 171 | 0 | 0 | 0 | 0 | 0 | 0 | 0 | 0 | 0 | 0 | 2 | 0 | 0 | 2 | 0 | 1 | 0 | 0 | 0 |
| 2 | 0 | 93 | 0 | 0 | 0 | 0 | 0 | 0 | 0 | 0 | 0 | 0 | 0 | 0 | 0 | 0 | 0 | 0 | 0 | 0 |
| 3 | 0 | 0 | 213 | 2 | 0 | 0 | 2 | 0 | 0 | 2 | 0 | 3 | 2 | 0 | 0 | 0 | 0 | 0 | 1 | 1 |
| 4 | 0 | 0 | 0 | 140 | 0 | 0 | 0 | 0 | 0 | 0 | 0 | 0 | 0 | 0 | 0 | 0 | 0 | 0 | 0 | 0 |
| 5 | 0 | 0 | 0 | 0 | 131 | 0 | 0 | 0 | 0 | 0 | 0 | 0 | 0 | 0 | 0 | 0 | 0 | 0 | 0 | 0 |
| 6 | 0 | 0 | 0 | 0 | 0 | 135 | 0 | 0 | 0 | 0 | 0 | 0 | 0 | 0 | 0 | 0 | 0 | 0 | 0 | 0 |
| 7 | 0 | 0 | 0 | 0 | 2 | 0 | 226 | 0 | 0 | 0 | 0 | 0 | 0 | 0 | 0 | 0 | 0 | 0 | 0 | 0 |
| 8 | 0 | 0 | 0 | 0 | 0 | 0 | 0 | 114 | 0 | 2 | 0 | 0 | 0 | 0 | 0 | 0 | 0 | 0 | 0 | 2 |
| 9 | 0 | 0 | 0 | 0 | 0 | 0 | 0 | 0 | 132 | 0 | 0 | 0 | 0 | 0 | 0 | 0 | 0 | 0 | 0 | 0 |
| 10 | 0 | 1 | 2 | 0 | 0 | 0 | 3 | 0 | 0 | 128 | 0 | 0 | 0 | 0 | 0 | 0 | 0 | 0 | 0 | 0 |
| 11 | 0 | 0 | 0 | 0 | 0 | 0 | 0 | 0 | 0 | 0 | 92 | 0 | 0 | 0 | 0 | 0 | 0 | 0 | 0 | 0 |
| 12 | 0 | 0 | 0 | 0 | 0 | 0 | 0 | 0 | 0 | 2 | 0 | 88 | 0 | 0 | 0 | 0 | 0 | 0 | 0 | 0 |
| 13 | 0 | 0 | 0 | 0 | 0 | 0 | 0 | 0 | 0 | 0 | 0 | 0 | 172 | 0 | 0 | 0 | 0 | 0 | 0 | 0 |
| 14 | 0 | 0 | 0 | 0 | 0 | 0 | 4 | 0 | 0 | 0 | 0 | 0 | 4 | 106 | 0 | 0 | 0 | 0 | 0 | 0 |
| 15 | 0 | 0 | 0 | 0 | 0 | 0 | 0 | 0 | 0 | 0 | 0 | 0 | 0 | 0 | 152 | 0 | 0 | 0 | 0 | 0 |
| 16 | 0 | 0 | 0 | 0 | 0 | 0 | 0 | 0 | 0 | 2 | 0 | 2 | 0 | 0 | 0 | 94 | 0 | 0 | 0 | 0 |
| 17 | 0 | 0 | 0 | 0 | 0 | 0 | 0 | 0 | 0 | 0 | 0 | 0 | 0 | 0 | 0 | 0 | 140 | 0 | 0 | 0 |
| 18 | 0 | 0 | 0 | 0 | 0 | 0 | 0 | 0 | 0 | 0 | 0 | 0 | 0 | 0 | 0 | 0 | 0 | 136 | 0 | 0 |
| 19 | 0 | 0 | 0 | 0 | 0 | 0 | 0 | 0 | 0 | 0 | 0 | 0 | 0 | 0 | 0 | 0 | 0 | 0 | 128 | 0 |
| 20 | 0 | 0 | 0 | 0 | 0 | 0 | 6 | 0 | 0 | 2 | 0 | 0 | 0 | 2 | 0 | 0 | 0 | 0 | 0 | 108 |

Table 5. Confusion-matrix CNN+BLSTM model

| CNN+B | 1 | 2 | 3 | 4 | 5 | 6 | 7 | 8 | 9 | 10 | 11 | 12 | 13 | 14 | 15 | 16 | 17 | 18 | 19 | 20 |
|---|---|---|---|---|---|---|---|---|---|---|---|---|---|---|---|---|---|---|---|---|
| 1 | 174 | 0 | 0 | 0 | 0 | 0 | 0 | 0 | 0 | 0 | 0 | 0 | 0 | 2 | 0 | 0 | 0 | 0 | 0 | 0 |
| 2 | 0 | 93 | 0 | 0 | 0 | 0 | 0 | 0 | 0 | 0 | 0 | 0 | 0 | 0 | 0 | 0 | 0 | 0 | 0 | 0 |
| 3 | 2 | 2 | 218 | 2 | 0 | 0 | 0 | 0 | 0 | 0 | 0 | 2 | 0 | 0 | 0 | 0 | 0 | 0 | 0 | 0 |
| 4 | 0 | 0 | 0 | 140 | 0 | 0 | 0 | 0 | 0 | 0 | 0 | 0 | 0 | 0 | 0 | 0 | 0 | 0 | 0 | 0 |
| 5 | 0 | 0 | 0 | 0 | 131 | 0 | 0 | 0 | 0 | 0 | 0 | 0 | 0 | 0 | 0 | 0 | 0 | 0 | 0 | 0 |
| 6 | 3 | 0 | 0 | 0 | 0 | 132 | 0 | 0 | 0 | 0 | 0 | 0 | 0 | 0 | 0 | 0 | 0 | 0 | 0 | 0 |
| 7 | 0 | 0 | 0 | 0 | 0 | 0 | 228 | 0 | 0 | 0 | 0 | 0 | 0 | 0 | 0 | 0 | 0 | 0 | 0 | 0 |
| 8 | 0 | 0 | 0 | 0 | 0 | 0 | 0 | 116 | 2 | 0 | 0 | 0 | 0 | 0 | 0 | 0 | 0 | 0 | 0 | 0 |

| | 1 | 2 | 3 | 4 | 5 | 6 | 7 | 8 | 9 | 10 | 11 | 12 | 13 | 14 | 15 | 16 | 17 | 18 | 19 | 20 |
|---|---|---|---|---|---|---|---|---|---|---|---|---|---|---|---|---|---|---|---|---|
| 9 | 0 | 0 | 0 | 0 | 0 | 0 | 0 | 0 | 132 | 0 | 0 | 0 | 0 | 0 | 0 | 0 | 0 | 0 | 0 | 0 |
| 10 | 0 | 0 | 0 | 0 | 0 | 0 | 0 | 0 | 0 | 130 | 0 | 0 | 0 | 0 | 0 | 2 | 0 | 0 | 0 | 2 |
| 11 | 0 | 0 | 0 | 0 | 0 | 0 | 0 | 0 | 0 | 0 | 92 | 0 | 0 | 0 | 0 | 0 | 0 | 0 | 0 | 0 |
| 12 | 0 | 0 | 0 | 0 | 0 | 0 | 0 | 0 | 0 | 0 | 0 | 88 | 0 | 0 | 0 | 2 | 0 | 0 | 0 | 0 |
| 13 | 0 | 0 | 0 | 0 | 0 | 0 | 0 | 2 | 0 | 0 | 0 | 0 | 170 | 0 | 0 | 0 | 0 | 0 | 0 | 0 |
| 14 | 0 | 2 | 0 | 0 | 0 | 0 | 0 | 4 | 0 | 0 | 0 | 0 | 0 | 108 | 0 | 0 | 0 | 0 | 0 | 0 |
| 15 | 0 | 0 | 0 | 0 | 0 | 0 | 0 | 0 | 0 | 0 | 0 | 0 | 0 | 0 | 152 | 0 | 0 | 0 | 0 | 0 |
| 16 | 0 | 0 | 0 | 0 | 0 | 0 | 2 | 0 | 0 | 2 | 0 | 0 | 0 | 0 | 0 | 92 | 0 | 0 | 0 | 2 |
| 17 | 0 | 0 | 0 | 0 | 0 | 0 | 0 | 0 | 0 | 0 | 0 | 0 | 0 | 0 | 0 | 0 | 140 | 0 | 0 | 0 |
| 18 | 0 | 0 | 0 | 0 | 0 | 0 | 0 | 0 | 0 | 0 | 0 | 0 | 0 | 0 | 0 | 0 | 0 | 136 | 0 | 0 |
| 19 | 0 | 0 | 0 | 0 | 0 | 0 | 0 | 0 | 0 | 0 | 0 | 0 | 0 | 0 | 0 | 0 | 0 | 0 | 128 | 0 |
| 20 | 2 | 0 | 2 | 0 | 0 | 0 | 0 | 2 | 0 | 0 | 0 | 0 | 0 | 0 | 0 | 2 | 0 | 2 | 0 | 108 |

Table 6. Confusion-matrix CNN+BLSTM model

| CNN+L | 1 | 2 | 3 | 4 | 5 | 6 | 7 | 8 | 9 | 10 | 11 | 12 | 13 | 14 | 15 | 16 | 17 | 18 | 19 | 20 |
|---|---|---|---|---|---|---|---|---|---|---|---|---|---|---|---|---|---|---|---|---|
| 1 | 172 | 0 | 2 | 0 | 0 | 0 | 0 | 0 | 0 | 0 | 0 | 0 | 0 | 0 | 2 | 0 | 0 | 0 | 0 | 0 |
| 2 | 0 | 93 | 0 | 0 | 0 | 0 | 0 | 0 | 0 | 0 | 0 | 0 | 0 | 0 | 0 | 0 | 0 | 0 | 0 | 0 |
| 3 | 2 | 0 | 217 | 0 | 0 | 0 | 0 | 2 | 0 | 0 | 0 | 1 | 0 | 0 | 2 | 0 | 0 | 0 | 0 | 2 |
| 4 | 0 | 0 | 0 | 138 | 1 | 0 | 0 | 0 | 0 | 0 | 0 | 0 | 0 | 0 | 0 | 0 | 0 | 0 | 1 | 0 |
| 5 | 0 | 0 | 0 | 0 | 131 | 0 | 0 | 0 | 0 | 0 | 0 | 0 | 0 | 0 | 0 | 0 | 0 | 0 | 0 | 0 |
| 6 | 0 | 0 | 0 | 0 | 0 | 135 | 0 | 0 | 0 | 0 | 0 | 0 | 0 | 0 | 0 | 0 | 0 | 0 | 0 | 0 |
| 7 | 0 | 0 | 0 | 0 | 0 | 0 | 228 | 0 | 0 | 0 | 0 | 0 | 0 | 0 | 0 | 0 | 0 | 0 | 0 | 0 |
| 8 | 0 | 0 | 0 | 0 | 0 | 0 | 0 | 114 | 0 | 0 | 0 | 0 | 2 | 0 | 0 | 2 | 0 | 0 | 0 | 0 |
| 9 | 0 | 0 | 0 | 0 | 0 | 0 | 0 | 0 | 132 | 0 | 0 | 0 | 0 | 0 | 0 | 0 | 0 | 0 | 0 | 0 |
| 10 | 0 | 0 | 0 | 0 | 0 | 0 | 0 | 2 | 0 | 125 | 0 | 2 | 0 | 2 | 0 | 2 | 0 | 0 | 1 | 0 |
| 11 | 0 | 0 | 0 | 0 | 0 | 0 | 0 | 0 | 0 | 0 | 92 | 0 | 0 | 0 | 0 | 0 | 0 | 0 | 0 | 0 |
| 12 | 0 | 0 | 2 | 0 | 0 | 0 | 0 | 0 | 0 | 0 | 0 | 86 | 0 | 0 | 0 | 2 | 0 | 0 | 0 | 0 |
| 13 | 0 | 0 | 2 | 0 | 0 | 2 | 0 | 0 | 0 | 0 | 0 | 0 | 166 | 2 | 0 | 0 | 0 | 0 | 0 | 0 |
| 14 | 0 | 0 | 0 | 0 | 0 | 0 | 0 | 0 | 0 | 0 | 0 | 0 | 0 | 114 | 0 | 0 | 0 | 0 | 0 | 0 |
| 15 | 0 | 0 | 0 | 0 | 0 | 0 | 0 | 0 | 0 | 0 | 0 | 0 | 0 | 0 | 152 | 0 | 0 | 0 | 0 | 0 |
| 16 | 0 | 0 | 0 | 0 | 0 | 0 | 0 | 0 | 2 | 0 | 0 | 2 | 0 | 92 | 0 | 2 | 0 | 0 | 0 | 0 |
| 17 | 0 | 0 | 0 | 0 | 0 | 0 | 0 | 0 | 0 | 0 | 0 | 0 | 0 | 0 | 0 | 140 | 0 | 0 | 0 | 0 |
| 18 | 0 | 0 | 0 | 0 | 0 | 0 | 0 | 0 | 0 | 0 | 0 | 0 | 0 | 0 | 0 | 0 | 136 | 0 | 0 | 0 |
| 19 | 0 | 0 | 0 | 0 | 0 | 0 | 0 | 0 | 0 | 0 | 0 | 0 | 0 | 0 | 0 | 0 | 0 | 0 | 128 | 0 |
| 20 | 2 | 0 | 0 | 2 | 0 | 0 | 0 | 0 | 0 | 0 | 0 | 0 | 0 | 0 | 0 | 0 | 0 | 0 | 2 | 112 |

Table 7. Comparison of the current study with other state of the art methods

| Paper | Technique | Language | Dataset | Recognition Ratio |
|---|---|---|---|---|
| [17] | MFCC+CNN | Pashto | Pashto digit database | 84.17% |
| [19] | Feed-forward TDNN | English | Tedlium2 | Word error rate (WER) of 7.6 in recognizing speech<br>Accuracy of 71.7% on predicting emotion from speech |
| [16] | MFCC+ANN | Arabic | Pronunciation of 3 Arabic letters | 92.42% |
| [6] | Deep Denoising Autoencoder + Inverse Filtering | Serbian | Whi-Spe | 92.81% |
| [18] | CNN with attention | English | Google Speech Commands | 91.4%<br>(94.5% for the 20-commands recognition task) |
| [14] | MFCC + Linear Predictive Coding (LPC) | Kannada | Kannada vowels dataset | 40% |
| [11] | CNN | English | Switchboard-1 | 97.0% |
| [10] | Connectionist | English | TIMIT | 18.2% phoneme error rate on the core test |

| | Temporal Classification (CTC) +CNN | | | set |
|---|---|---|---|---|
| [8] | Hybrid convolutional neural network (HCNN) | English | English isolated word speech corpus along with TVs | - |
| [15] | MFCC+ deep belief network (DBN) LSTM, BLSTM, DLSTM, and DBLSTM | Persian | Farsdat | HMM: 75.2%, LSTM: 77% LSTM-DBN: 78%, BLSTM: 79.3% Karel-DNN: 80.2%, DLSTM: 80.3% DBLSTM: 82.9%, DBLSTM-DNN: 83.2% |
| This study | MFCC, CNN, LSTM, MLP, CNN+BLTSM, CNN+LSTM | Dari | Our built in Dari speech dataset | Average accuracies: CNN: 98.037 %, LSTM: 97.712 %, MLP: 97.02% CNN+LSTM: 98.184 %, CNN+BLSTM: 98.365% |